\def\slaninafigdir{./}
\newcommand{\slaninafigafter}[1]{}
\newcommand{\slaninafigure}[2]{
\begin{figure}[t]
  \centering
  \vspace*{70mm}
  \includegraphics{\slaninafigdir#1.ps}
  \caption{#2}
  \label{fig:#1}
\end{figure}
}
\newcommand{\slaninafigurebig}[2]{
\begin{figure}[t]
  \centering
  \vspace*{125mm}
  \includegraphics{\slaninafigdir#1.ps}
  \caption{#2}
  \label{fig:#1}
\end{figure}
}
\documentclass{elsart}
\usepackage{latexsym}
\begin{document}
\begin{frontmatter}
\title{Harms and benefits from social imitation 
}
\author{Franti\v{s}ek Slanina}
\address{
        Institute of Physics,
	Academy of Sciences of the Czech Republic,\\
	Na~Slovance~2, CZ-18221~Praha,
	Czech Republic\\
        e-mail: slanina@fzu.cz
}
\begin{abstract}

We study the role of imitation within 
a model of economics with adaptive agents. The basic ingredients are
those of the 
Minority Game. We add the
possibility of local information exchange and imitation of the
neighbour's strategy. Imitators should pay a fee to the
imitated. 
Connected groups are formed,
which act as if they were single players.
Coherent spatial areas of rich and poor agents result, leading to the
decrease of local social tensions.
Size and stability of these areas depends on the parameters of the model. 
Global performance measured by the attendance
volatility is optimised at certain value of the imitation probability.
The social tensions are suppressed for large imitation probability,
but due to the price paid by the imitators 
the requirements of high global effectivity and low
social tensions are in conflict, as well as the requirements of low
global and low local wealth differences.
\end{abstract}


\end{frontmatter}
\section{Introduction}
The Minority Game introduced by Challet and Zhang 
\cite{cha_zha_97,cha_zha_98}
following the earlier ideas of B. Arthur \cite{arthur_94} became in recent 
years a playing ground for studying various aspects of the economic systems.

In the Minority Game (MG) we have $N$ players who choose repeatedly
between two options and compete to be in the minority 
group. This is the idealisation of various situations, where the
competition for limited resources leads to intrinsic frustration. One
can think for example of cars choosing between two alternative routes
or a speculator who tries to earn money by buying and selling shares
in such a manner that 
the majority takes the opposite action than herself.

Let us recall some well-known facts about the MG. The players share a
public information, saying what were the outcomes  
of the game in past $M$ rounds. The players interact only through this
information. Therefore, the system has a ``mean-field'' character, in
the sense that no short-range interactions exist. 

The self-organization is achieved by allowing players to have several
strategies and choose among them the strategy which seems to be the best
one. This feature leads to decrease of the fluctuations of
attendance below its random coin-tossing value, thus increasing the
global effectivity of the system. It was found that the relevant
parameter is $\alpha=2^M/N$ and the maximum effectivity is reached for
$\alpha=\alpha_c\simeq 0.34$ 
\cite{cha_zha_98,sa_ma_ri_99,jo_ja_jo_che_kwo_hui_98} and the properties
of this phase transition are thoroughly studied using the methods 
developed in the theory of neural networks \cite{cha_ma_99,cha_ma_ze_00,dem_mar_00}.

More complete account of the current state of the standard MG and its 
ramifications is given in other contributions in these Proceedings \cite{thisvolume}.
We would like to stress especially
the attempts to go back to the economic motivations of MG and model
the market mechanisms 
\cite{cha_ma_zha_99,sla_zha_99,jo_ha_hu_zhe_99,cha_che_mar_zha_00}.

The observation that the crowded (low $\alpha$) phase exhibits low 
global effectivity bears an important hint. Indeed, if we start with the 
crowded phase, we can improve the performance by grouping the agents together.
This mechanism may bring about the condensation of individual
investors around consulting companies and investment funds, which is
the behaviour found in real life.

Indeed, an individual investor who sees that she is all the time
behind her neighbors may feel tempted to refrain from her own
initiative and transfer the burden of decisions to more successful
(more wealthy) individuals. That is what we will call imitation. 
The temptation for imitation in the population will be quantified by a
parameter $p\in[0,1]$. Of course, an agent, who is otherwise
prone to imitation, will {\it not} imitate, if she has larger wealth
and therefore is better off than her neighbors. So, there are be two
questions to be positively answered if the imitation is to occur: Has
the agent natural tendency to imitation? Has any of her neighbors
larger wealth?

It is also natural to suppose that the 
decision-maker, or the imitated individual, will use (or misuse)
her position to require a fee from those on which behalf her acts.
Therefore, the imitators will pay a commission $\varepsilon$ to the
imitated. As we will see, the value of the commission has important
consequences for the behavior of the agents.

We introduced recently \cite{slanina_00} the possibility of local
interactions 
into the standard MG.
In this contribution we further analyse the properties of
social structures emerging from the local information exchange.
When doing so we 
go beyond the mean-field
character of the usual MG. 
Related works were already done, either assuming that the global
information is fully replaced by a local one \cite{kal_schu_bri_00}
or using the MG scheme for evolving the Kauffmans's Boolean networks
\cite{kauffman_90a} to the critical state \cite{pa_bas_99}.

In our variant of the MG the local information  is used to enable the
players to decide, whether they want to use their own strategies or 
imitate their neighbours. 
Indeed, it is quite
common that people do not invest individually, but rely on an advice
from specialised agencies, or simply follow the trend they perceive in
their information neighbourhood. In so doing, the individuals coalesce
into groups, which act as single players. In the framework of 
Minority Game, we will study the social structure induced by the
occurrence of these groups. It should be expected that this will lead 
to increase in the global performance in the crowded (small $\alpha$)
phase. This is indeed confirmed by the simulations.

\section{Minority Game on a chain with allowed imitation}
We introduce the possibility of local information exchange in our
variant of the Minority Game. In analogy to the metabolic pathways 
in living organisms we can imagine a kind of ``information
metabolism'' in work within the economic system. Information flow 
along the edges of certain information network. 
The study of the geometry of graphs describing these 
information networks is now a
scientific field on its own
\cite{wa_stro_98,bar_alb_99,sla_ko_00}. Within the framework of MG a
linear chain \cite{kal_schu_bri_00} and random network with 
fixed connectivity $K$ 
\cite{pa_bas_99} was
already investigated in different contexts.
 
Here we take the simplest possible choice of a linear chain with
one-directional nearest-neighbour connections. Each player can obtain
the information only from her left-hand neighbour, namely 
about her neighbour's wealth.

There will be two conditions needed for a player to imitate her
neighbour. First, the player should have internal disposition for being
an imitator. We simplify the variety of risk-aversion levels by
postulating only two types of players. 
Each player has a label $\tilde{l}\in\{{\rm 1},{\rm 0}\}$ indicating,
whether the player is a potential imitator ($\tilde{l}=1$) or
always a leader ($\tilde{l}=0$). 
 At the beginning we take each of the players and
attribute her the label 1 with probability $p$ and label 0 with
probability $1-p$. We also allow swapping between the two types of
behaviour, 
 at a constant rate. The labels can change at each step with probabilities
$p_1$ ($1\to 0$) and $p_2$ ($0\to 1$). We choose always $p=p_2/(p_1+p_2)$, 
so that the average density of potential imitators does not 
change in time.

The second condition for the player of the type 1 to actually imitate
in the current step is that her neighbour has larger accumulated wealth
than the player itself. We suppose that the player does not know what
are the strategies of her neighbour, but if she observes that the
neighbour's behaviour is more profitable than her own strategy, she
relegates the decision to the neighbour and takes the same action.
The player of the type 0 will never imitate. Therefore, she will always
look only at her $S$ strategies and choose the best estimate from
them. 

The above rules are formalized as follows.  We have an odd number $N$ of 
players. Each player has $S=2$
strategies, denoted $s_j\in\{1,2\}$. 
The two possible actions a player can take are $0$ and $1$. The
winning action is $1$ if most players took $0$ and vice versa. The
members of teh winning side receive 1 point, the loosing side 0 points.
The players know  the last $M$ outcomes of the game. This information
is arranged into the $M$-bit string $\mu\in\{0,1\}^M$. The strategies
are tables 
attributing to each of $2^M$ possible strings $\mu$ the action
$a_{j,s_j}^\mu$ the
player $j$ takes, if she chooses the strategy $s_j$. The scores
$U_{j,s}$ of the strategies are updated according 
to the minority rule
\begin{equation}
U_{j,s}(t+1)=U_{j,s}(t)+1-\delta(a_{j,s}^{\mu(t)}-\theta(\sum_i a_i(t) - N/2)) 
\end{equation}
where $a_j(t)$ is the action the player $j$ takes at time $t$. 

 The potential imitators 
will copy the action from their more successful neighbours.
Let $W_j$ be the wealth of the $j$-th player and the variables $l_j$
describe   the actual state of imitation, in analogy with the labels
$\tilde{l}_j$ describing potential state of imitation. We can write
$l_j =\tilde{l}_j\,\theta(W_{j-1}-W_j)$, with $\theta(x)=1$ for $x>0$
and 0 otherwise. The actions of the players are
\begin{equation}
a_j = l_j a_{j-1} + (1-l_j)a_{j,s_{\rm M}}\;\; .
\end{equation}

We also suppose that the imitation is not for free. The player who
imitates passes a small fraction $\varepsilon$ of its wealth increase to
the imitated player. This rule accounts for the price of
information. Then, we update the
wealth of players iteratively,
\begin{equation}
\Delta W_{j}(t) = 
(1-\varepsilon l_j)(\varepsilon l_{j+1}\,\Delta W_{j+1}(t)+1-\delta(
a_{j}-\theta( \sum_i a_i(t) - \frac{N}{2}))
\end{equation}
where $\Delta W_j(t) = W_{j}(t+1) - W_{j}(t)$.
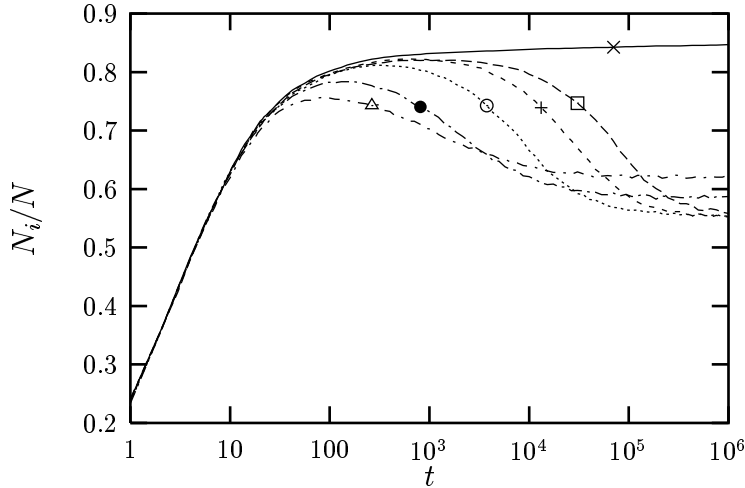
\slaninafigure{imi-vs-time-95-05}{Time dependence of the fraction of imitators,
for $N=1001$, $M=6$, $S=2$, and $p=0.95$, averaged over 10 independent runs. 
Different curves (marked by symbols)
correspond to different probability 
$p_1=0$ ($\times$), 
$5\cdot 10^{-6}$ ($\Box$), 
$1.5\cdot 10^{-5}$ 
(%
$+$%
)
$5\cdot 10^{-5}$ ($\odot$), 
$5\cdot 10^{-4}$ ($\bullet$), 
$5\cdot 10^{-3}$ ($\triangle$). 
}

\section{Imitation structures}

In our simulations we observe that the time evolution of the 
number of actually
imitating players, $N_i=\sum_j l_j$, depends on $p_1$.
The time dependence of
the fraction of imitators $N_i/N$ for several values of $p_1$ is shown in
Fig. \ref{fig:imi-vs-time-95-05}.
For $p_1=0$ it 
increases monotonously until 
saturation, while for $p_1 \ne 0$ it grows toward a local maximum and then
decreases and saturates at a value weakly dependent on $p_1$, but 
significantly below the $p_1=0$ value.

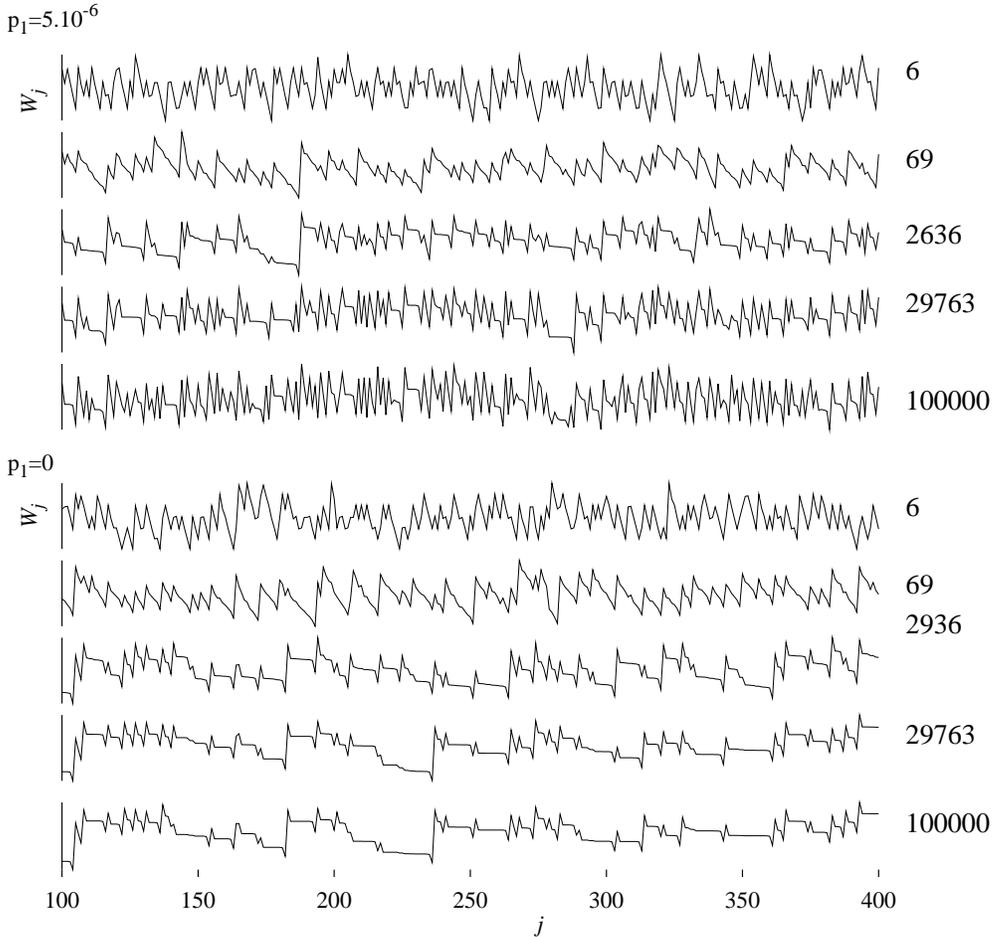
\slaninafigurebig{structure-both}{Example of the evolution of the 
distribution of wealth among
  players, for $N=1001$, $M=6$, $S=2$, $\varepsilon=0.05$, and $p=0.95$.
The upper 5 curves correspond to $p_1=5\cdot 10^{-6}$, while the 
lower 5 curves have $p_1=0$.
  The time
  step at which the 
  snapshot is taken is indicated on the right. For each time, the
  vertical axis indicates the wealth $W_j$ of the $j$-th player.}
An example of the time evolution of the spatial wealth
distribution is given in Fig. \ref{fig:structure-both}
for $p=0.95$ and two values of $p_1=5\cdot 10^{-6}$ and $p_1=0$.
The initially random 
distribution of wealth among players changes qualitatively during the
evolution of the system. Coherent groups of poor and wealthy players
are formed. Again, the situation is qualitatively different if we allow 
the players to switch between potential imitator and leaders.
We have shown in the previous work \cite{slanina_00} that for $p_1=0$ 
the poor groups
persist forever. We can see the same behaviour also in 
Fig. \ref{fig:structure-both} for $p_1=0$. On the other hand,
for $p_1\ne 0$ we observe that large poor groups are unstable and split 
again into smaller clusters.
This leads to lowering of the global wealth differences, as will be 
analysed in the next section.

\section{Globally uniform wealth versus small social tensions}

The time averaged attendance fluctuations 
$\sigma^2=\langle (A-N/2)^2\rangle$ measure the distance from the
global optimum. The global effectivity is higher for smaller
$\sigma^2$. 
We investigated the influence of the imitation on the global
effectivity.

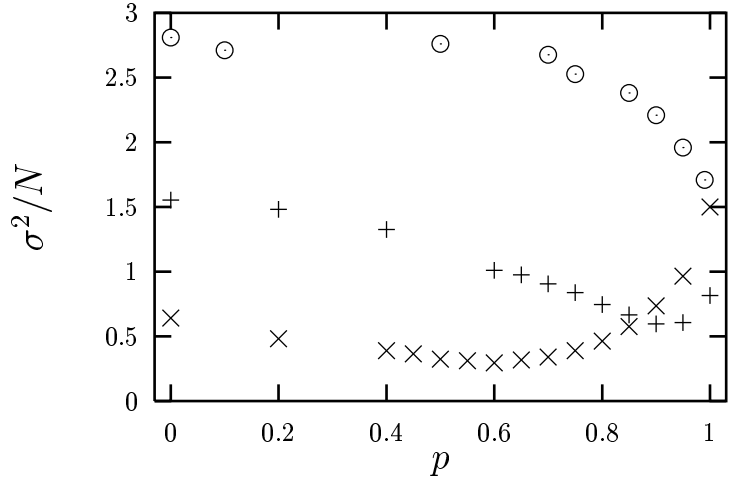
\slaninafigure{sig-all-all}{Dependence of the attendance fluctuations
 on the imitation probability for $p_1=0$.
 The number of players is
 $N=1001$ and 
 memory length $M=5$  ($\odot$), $M=6$ ($+$), and $M=7$ ($\times$).}
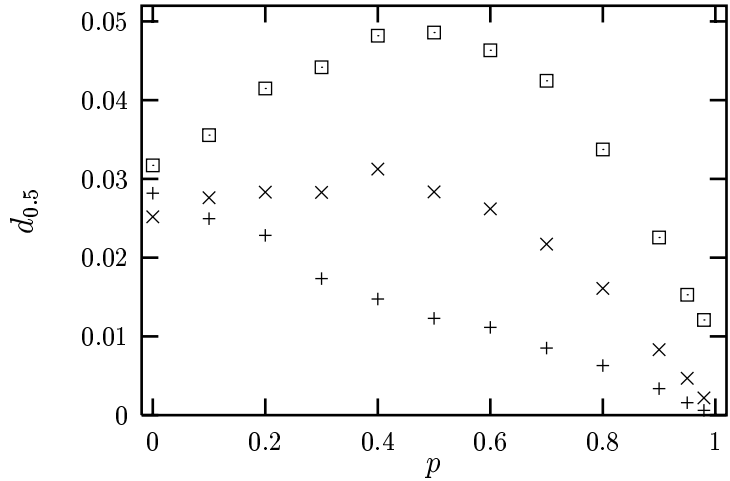
\slaninafigure{tensions-all}{Relative local tension for $N=1001$, $M=6$, 
$p_1=0$ measured
 by utility  function   $(\Delta W)^{1/2}$
 for commission $\varepsilon=0.05$ ($\Box$), $0.03$ ($\times$), 
and $0.01$ ($+$).}
We found that in the crowded
phase the system becomes more efficient if imitation is allowed
($p>0$), but there is a local minimum in the dependence of $\sigma^2/N$
on $p$, indicating that there is an optimal level of imitation,
beyond which the system starts to perform worse. The results for
$N=1001$ are shown in Fig.  
 \ref{fig:sig-all-all}. We can see that the minimum occurs at smaller
values for larger $M$. We can also observe that for longer memories
($M=7$ in our case) the value of the fluctuations for $p=1$ is
significantly above the value without imitation ($p=0$), while the
value at the minimum still lies below the $p=0$ value. This implies
that moderate imitation can be beneficiary, while exaggerated one can
be harmful.

The increase of spatial coherence by creation of poor and wealthy groups 
can result in decrease
of local social tension. To quantify it, we introduce
 a kind of
``utility function'' \cite{merton_90} $U(\Delta W)$, which indicates,
how much the wealth difference $\Delta W$ is subjectively
perceived.
We will use the utility function in the form
$U(x)=x^{1/2}$. Then, the average measure of the local
social tension is
\begin{equation}
d_{0.5}=\frac{1}{\langle
W\rangle}\left(\sum_{j=1}^{N-1} |W_j-W_{j+1}|^{1/2} \right)^{2} 
\end{equation}
where we denoted the average wealth 
$\langle W\rangle=\frac{1}{N}\sum_{j=1}^{N} W_j$.

The stationary values of the tension
 for various values of the commission $\varepsilon$ are shown in 
Fig. \ref{fig:tensions-all}, for $p_1=0$. An important feature 
of the $p$-dependence is the maximum at certain imitation probability.
The maximum becomes more pronounced for larger commission $\varepsilon$, 
while for $\varepsilon=0.01$ it disappears.

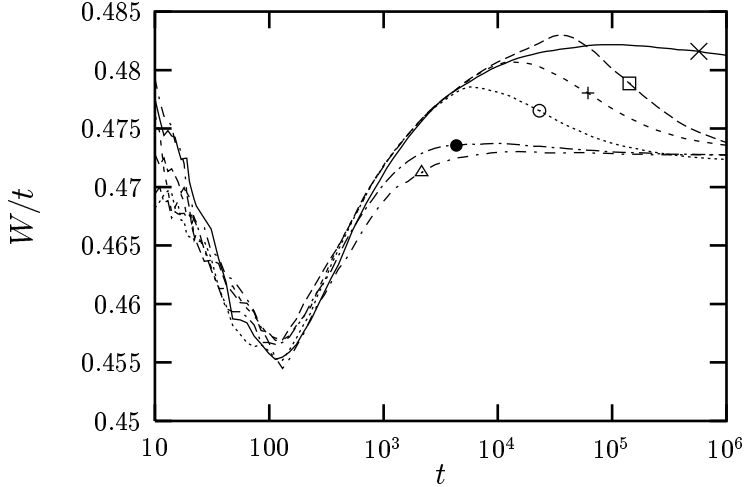
\slaninafigure{wealth-growth-long}{Growth of the average wealth of
 agents, for $N=1001$, $M=6$, $S=2$, $p=0.95$,
sample-averaged over 10 independent runs. 
Different curves (marked by symbols)
correspond to different probability 
$p_1=0$ ($\times$), 
$5\cdot 10^{-6}$ ($\Box$), 
$1.5\cdot 10^{-5}$ 
(%
$+$%
)
$5\cdot 10^{-5}$ ($\odot$), 
$5\cdot 10^{-4}$ ($\bullet$), 
$5\cdot 10^{-3}$ ($\triangle$).}

This observation has an important consequence. Imagine, we are 
social experimentalists starting with a system with no information
exchange and no imitation. Let us try to lower the social tensions
by gradually encouraging the people to buy information from the neighbours
and imitate each other. If the cost of the information ($\varepsilon$)
is too high, this social strategy would fail, because small increase
in imitation would {\it enhance} the social tension. Lower social
tension must have been achieved by a macroscopic change in the social
behaviour: by jumping over the maximum in the function $d_{0.5}(p)$.
This may serve as a toy example of how too greedy environment (too
costly information) can prevent the system to find a global optimum.

By comparing the Figs.  \ref{fig:sig-all-all} and
\ref{fig:tensions-all} 
we can also see that for high $\varepsilon$ 
optimal performance (minimum $\sigma^2/N$) can be 
close to maximum in social tensions. Therefore, in greedy environment
the requirements of effectivity and social peace are in conflict.

The 
Fig. \ref{fig:wealth-growth-long}
shows the growth rate of the average wealth for several values of the
switching probability $p_1$. We can see that the growth rate converges
to a constant value, which is higher for $p_1=0$ and nearly
independent of $p_1$ for $p_1\ne 1$. In all cases we confirm that 
the average wealth grows linearly with time.

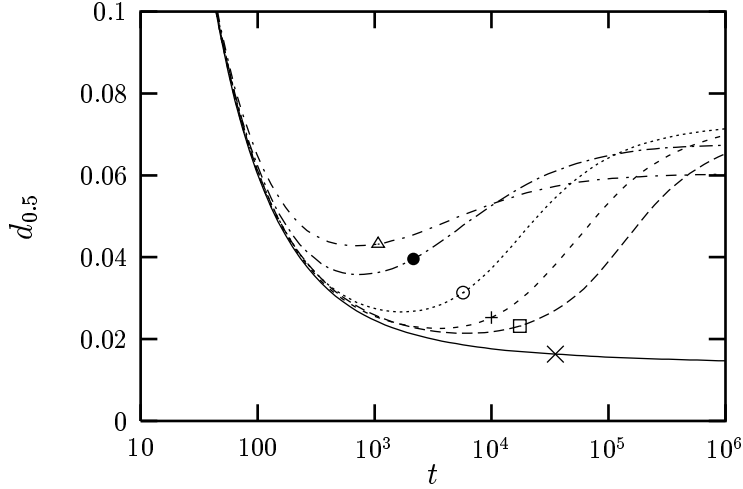
\slaninafigure{tensions-long}{Time evolution of the local tensions,
for $N=1001$, $M=6$, $S=2$, $p=0.95$,  
sample-averaged over 10 independent runs. 
Different curves (marked by symbols)
correspond to different probability 
$p_1=0$ ($\times$), 
$5\cdot 10^{-6}$ ($\Box$), 
$1.5\cdot 10^{-5}$ 
(%
$+$%
)
$5\cdot 10^{-5}$ ($\odot$), 
$5\cdot 10^{-4}$ ($\bullet$), 
$5\cdot 10^{-3}$ ($\triangle$).}
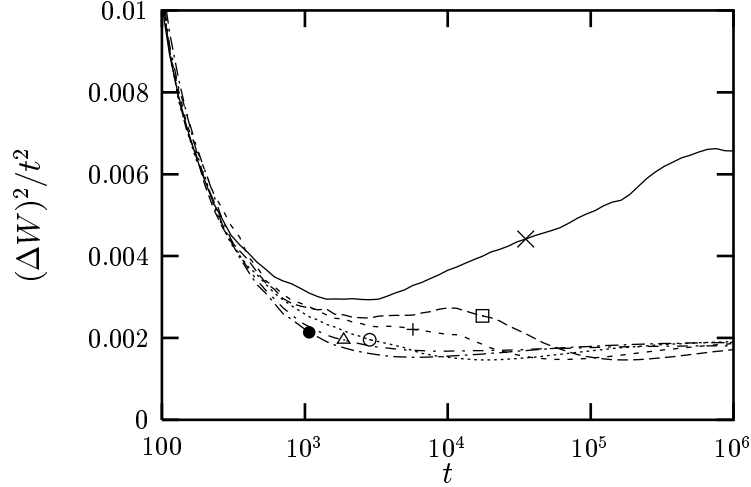
\slaninafigure{wealth-width-growth-long}{Time evolution of the wealth
dispersion, for $N=1001$, $M=6$, $S=2$, $p=0.95$, averaged over 10
independent runs.  
Different curves (marked by symbols)
correspond to different probability 
$p_1=0$ ($\times$), 
$5\cdot 10^{-6}$ ($\Box$), 
$1.5\cdot 10^{-5}$ 
(%
$+$%
)
$5\cdot 10^{-5}$ ($\odot$), 
$5\cdot 10^{-4}$ ($\bullet$), 
$5\cdot 10^{-3}$ ($\triangle$).}
In the 
Fig. \ref{fig:tensions-long}
we can see the time evolution of the local tensions
for several values of the switching probability $p_1$. We observe that
the switching enhances the local tensions. On the other hand, in Fig.
 \ref{fig:wealth-width-growth-long}
we can see the time dependence of the growth rate in the global
wealth dispersion, $\langle W^2\rangle-\langle W\rangle^2$ (by angle
brackets we denote
the average over all players). There is a clear difference between the
cases of $p_1=0$, where the wealth dispersion grows much more rapidly
than $t^2$
 and $p_1\ne 0$, where the dispersion
grows as $t^2$, at a rate nearly independent of $p_1$.

This means that if we allow switching between potential imitation and 
leader states, the wealth distribution only re-scales linearly in time
(This observation together with the linear growth if the average
wealth
suggests that the 
probability density at time $t$ converge as $P(W,t)=\Phi(W/t)$
where the function $\Phi(x)$ does not depend on time).
On the contrary, if we forbid the switching, the poor imitators are frozen
forever in their poverty and in the wealth distribution the rich and
poor diverge steadily.

However, recalling the discussion of the Figs.
\ref{fig:tensions-long} and 
 \ref{fig:wealth-growth-long}
we can see that the requirement of low global wealth dispersion
(a ``just'' world, achieved by enabling the poor imitators switch to
leaders and thus become richer) deteriorates both global efficiency
(measured now by the wealth growth rate) and, more surprisingly, the
local social tensions.

\section{Conclusions}
We investigated the creation of rich and poor spatial domains due to
local information exchange, within the framework of the Minority
Game (MG). Coherent spatial areas of rich and poor agents emerge.
Several macroscopic conflicts of interest are observed in our model.

(1)
We found that the effect of imitation leads to increased
effectivity in the crowded phase of MG. The price paid for the
information
needed to imitation
leads to the conflict between effectivity and local social tensions.
High information cost also prevents the system from coming to
the state of lower social tensions by gradual increase of the 
imitation probability.

(2)
We allow for switching between imitation and non-imitation (leader) 
states. Such a switching makes the global wealth differences smaller, but
increases the local social tensions.

The creation of coherent areas of poor and rich agents leads to
decrease in the local social tensions, but only if $p$ is sufficiently
close to 1. The lowest value of the social tension is reached at
$p=1$, but for such a value the global effectivity is significantly
lower than its optimum value. Therefore, we observe a conflict of
local interests (maximisation of social tension) with global
performance (maximisation of attendance fluctuations).

\section*{Acknowledgments}
I wish to thank to Yi-Cheng Zhang for numerous useful discussions. 
This work was supported by the Grant Agency of the Czech Republic,
grant project No. 202/01/1091. I acknowledge the financial support 
from the University of Fribourg,
Switzerland, where part of this work was done.

%pagebreak

\end{document}